\documentclass[12pt,twoside]{article}
\usepackage{fleqn,espcrc1,epsfig}

\usepackage{graphicx}
\usepackage[figuresright]{rotating}

\newcommand{\AmS}{{\protect\the\textfont2
  A\kern-.1667em\lower.5ex\hbox{M}\kern-.125emS}}

\title{Laying the groundwork at the AGS: Recent results from experiment E895}

\author{
M.A.~Lisa\address[OSU]{\vspace{-3mm} \small Ohio State University, Columbus, Ohio},
N.N.~Ajitanand\address[SUNY]{\vspace{-3mm} \small State University of New York, Stony Brook},
J.M.~Alexander\addressmark[SUNY],
M.Anderson\address[UCD]{\vspace{-3mm} \small University of California, Davis, California},
D.~Best\address[LBL]{\vspace{-3mm} \small Lawrence Berkeley National Laboratory, Berkeley, California},
F.P.~Brady\addressmark[UCD],
T.~Case\addressmark[LBL],
W.~Caskey\addressmark[UCD],
D.~Cebra\addressmark[UCD],
J.L.~Chance\addressmark[UCD],
P.~Chung\addressmark[SUNY],
B.~Cole\address[COLUMBIA]{\vspace{-3mm} \small Columbia University, New York, New York},
K.~Crowe\addressmark[LBL],
A.C.~Das\addressmark[OSU],
J.E.~Draper\addressmark[UCD],
M.L.~Gilkes\addressmark[SUNY],
S.~Gushue\address[BNL]{\vspace{-3mm} \small Brookhaven National Laboratory, Upton, New York},
M.~Heffner\addressmark[UCD],
A.S.~Hirsch\address[PURDUE]{\vspace{-3mm} \small Purdue University, West Lafayette, Indiana},
E.L.~Hjort\addressmark[PURDUE],
L.~Huo\address[HARBIN]{\vspace{-3mm} \small Harbin Institute of Technology, Harbin, P.R.~China},
M.~Justice\address[KENT]{\vspace{-3mm} \small Kent State University, Kent, Ohio},
M.~Kaplan\address[CMU]{\vspace{-3mm} \small Carnegie Mellon University, Pittsburgh, Pennsylvania},
D.~Keane\addressmark[KENT],
J.C.~Kintner\address[STMARY]{\vspace{-3mm} \small St.~Mary's College, Moraga, California},
J.~Klay\addressmark[UCD],
D.~Krofcheck\address[NZ]{\vspace{-3mm} \small University of Auckland, Auckland, New Zealand},
R.A.~Lacey\addressmark[SUNY],
C.~Law\addressmark[SUNY],
J.~Lauret\addressmark[SUNY],
H.~Liu\addressmark[KENT],
Y.M.~Liu\addressmark[HARBIN],
R.L.~McGrath\addressmark[SUNY],
Z.~Milosevich\addressmark[CMU],
G.~Odyniec\addressmark[LBL],
D.L.~Olson\addressmark[LBL],
C. Pinkenburg\addressmark[SUNY]\addressmark[BNL],
S.~Panitkin\addressmark[KENT],
N.T.~Porile\addressmark[PURDUE],
G.~Rai\addressmark[LBL],
H.G.~Ritter\addressmark[LBL],
J.L.~Romero\addressmark[UCD],
R.~Scharenburg\addressmark[PURDUE],
L.S.~Schroeder\addressmark[LBL],
B.~Srivastava\addressmark[PURDUE],
N.T.B.~Stone\addressmark[LBL],
T.J.M.~Symons\addressmark[LBL],
S.~Wang\addressmark[KENT],
R.~Wells\addressmark[OSU]
J.~Whitfield\addressmark[CMU],
T.~Wienold\addressmark[LBL],
R.~Witt\addressmark[KENT],
L.~Wood\addressmark[UCD],
W.N.~Zhang\addressmark[HARBIN]
}

\begin{document}

\maketitle

\begin{abstract}

The E895 Collaboration at the Brookhaven AGS has performed
a systematic investigation of Au+Au collisions at 2-8 AGeV,
using a large-acceptance Time Projection Chamber.  In addition
to extensive measurements of particle flow, spectra, two-particle
interferometry, and strangeness production, we have performed novel
hybrid analyses, including azimuthally-sensitive pion HBT,
extraction of the six-dimensional pion phasespace density, and
a first measurement of the $\Lambda$-p correlation function.

\end{abstract}

\section{Introduction}
\label{sec-intro}

Modern relativistic heavy ion physics focuses largely on the creation and
study of the Quark Gluon Plasma (QGP), a state of matter in which partonic
degrees of freedom are relevant over large length scales;
potential signatures of QGP formation (see, e.g.~\cite{bass-gyulassy-qgp-sig})
are studied in collisions at AGS, SPS, and now RHIC.
However, it is now well-accepted that only through a simultaneous study of several
observables will a clear understanding of the collision history emerge.  Furthermore,
the most convincing evidence of ``new'' physics stems from a break in well-understood
(in terms of hadronic models) systematics in observables which have been measured over
a large range of conditions.

Using the EOS Time Projection Chamber (TPC)~\cite{rai-ieee} at the 
Brookhaven AGS, we have studied Au+Au collisions at bombarding energies of 2, 4, 6,
and 8~AGeV.  Models suggest that collisions at these energies reach the highest baryon
density~\cite{pang92,li96,dani98}, making such systems ideal to stringently test
our understanding of nuclear matter under extreme conditions, as well as to search for
possible QGP formation in the high-density limit~\cite{dani98,baym76}.

We have measured excitation functions in four primary areas: (1) proton elliptic and directed 
flow~\cite{E895-v2,E895-v1}, (2) intensity
interferometry~\cite{E895-HBT1,E895-HBT2,E895-Sergei,E895-QM99-Lisa},
(3) charged particle spectra~\cite{E895-Klay,E895-spectra}, and (4) strange particle 
production~\cite{E895-QM01-Pinkenburg}.  In addition to improving our understanding
of hot hadronic systems under extreme conditions, these systematics provide a valuable
baseline for measurements at higher energy.

Going further, we have combined the analyses in these areas to achieve novel
insights into the collision dynamics, and serve as an example of how to optimize
physics information extraction for data at RHIC.  In~\cite{E895-QM01-Pinkenburg,E895-LambdaFlow,E895-K0Flow},
strange particle reconstruction is correlated with proton flow to extract information
on in-medium potentials.  In this report, we focus primarily on combined analyses involving
two-particle correlations.

\vspace*{-3mm}
\section{Directed and Elliptic Flow}
\label{sec-flow}

Anisotropic flow refers to the momentum distribution of particles in azimuth about the
beam axis.  The reference azimuthal angle is the direction of the impact parameter vector ${\bf b}$,
which must be determined on an event-by-event basis; see~\cite{E895-v2} for details.

Two types of anisotropic flow are typically discussed~\cite{flow-review}.
(Positive) sideward flow refers to the preferential in-plane emission of particles
with $y>y_{cm}$ in the direction of ${\bf b}$, and those with $y<y_{cm}$ opposite ${\bf b}$,
while elliptical flow analyses focus on the difference between in- and out-of-plane emission.
In a simple picture in which the emission pattern can be described by an ellipsoid in momentum
space (in general, the situation is more complicated~\cite{csernai}), then, sideward flow
analyses measure the tilt angle of the momentum-space ellipsoid,
while
elliptical flow probes the ratio between its semi-minor axes.

E895 analyses of proton directed and elliptic flow have already been
reported~\cite{E895-v2,E895-v1,E895-QM99-Rai}.   There, we found that
AGS energies represent a transition region, in which
the elliptic flow
changes sign from out-of-plane (the ``squeeze-out'' signal observed at Bevalac/SIS energies)
to in-plane elliptic flow seen at SPS energies,
and directed flow reaches a maximum and falls off quickly with beam energy.
{\it Pion} directed flow is oriented in the direction opposite proton flow-- i.e.
the $\pi$ momentum-space ellipse tilts away from ${\bf b}$ for $p_z>0$~\cite{LHW00},
while elliptic flow changes sign at $E_{beam}\approx$4~AGeV, similar to the protons~\cite{caskeyPhD}.

Dynamic transport models reproduce the gross trends in flow measured by E895, but the
details are strongly sensitive to the underlying physics (e.g. the Equation of State)
assumed~\cite{E895-v2,E895-v1}; hence, these excitation functions provide a very stringent
test of our understanding of the dynamics of heavy ion collisions.  In particular,
no single model is currently able to reproduce the detailed flow excitation function
at AGS energies.

Clearly, fresh information on the physics of anisotropic flow is needed.
It is commonly assumed 
that these flow signals result from space-momentum
correlations generated in the collision.
By measuring azimuthally-sensitive
momentum distributions, therefore, we look at only half the story.
In the next section, we describe how two-pion correlations provide a
more complete picture.


\section{Pion Intensity Interferometry}
\label{sec-HBT}
\vspace*{-2mm}

Intensity interferometry (a.k.a. HBT) techniques have been used extensively to probe
the space-time structure of heavy ion collisions (see, e.g.~\cite{bauer_gelbke_pratt,WH99}).
Until recently, most HBT analyses implicitly assumed a cylindrically symmetric emission
region, strictly valid only for $|{\bf b}|=0$ collisions.
In this standard framework, E895 $\pi^-$ correlation systematics
as a function of beam energy, $m_T$, centrality, and rapidity, have already been
reported~\cite{E895-HBT1,E895-QM99-Lisa}.

In addition to the (momentum-space) anisotropic flow studies discussed above,
theoretical studies~\cite{nutcracker,heiselberg,frankfurt-hydro,W98,LHW00,VoloshinHBTflow}
suggest that {\it coordinate-space} anisotropies of the emitting source are equally interesting.
Recently, a formalism was proposed to use azimuthally-sensitive intensity interferometry,
correlated event-by-event with the reaction plane,
to extract this information experimentally~\cite{W98,LHW00}.
We have performed
the first experimental measurement of the full
coordinate-space anisotropies of a hot nuclear source, for pions emitted from Au+Au collisions
at 2-6 AGeV.
Full details may be found in~\cite{E895-HBT2}.

$\pi^-$ correlation functions were constructed in the Bertsch-Pratt
(``out-side-long'') decomposition as measured in
the Au+Au c.m. frame~\cite{BP}, with cuts
on the pair angle with respect to the reaction plane
$\phi=\angle({\bf K_\perp},{\bf b})$, where ${\bf K_\perp}=({\bf p_1} + {\bf p_2})_\perp$
is the total momentum of the pair perpendicular to the beam. 

The correlation functions for each of eight 45-degree-wide
$\phi$ bins, are fit with the standard Gaussian parameterization~\cite{WH99}
\begin{equation}
  \label{eq:extended-BP}
    C({\bf q},\phi) =
    1 + \lambda(\phi)\, 
    \exp\Bigl[- \sum_{i,j=o,s,l} q_i q_j R_{ij}^2(\phi) \Bigr]
    \,.
\end{equation}
In contrast to $\phi$-integrated analyses, all six HBT radii are relevant
here~\cite{WH99,W98,LHW00}.

\begin{figure}[ht]
\begin{minipage}[t]{80mm}
\epsfig{file=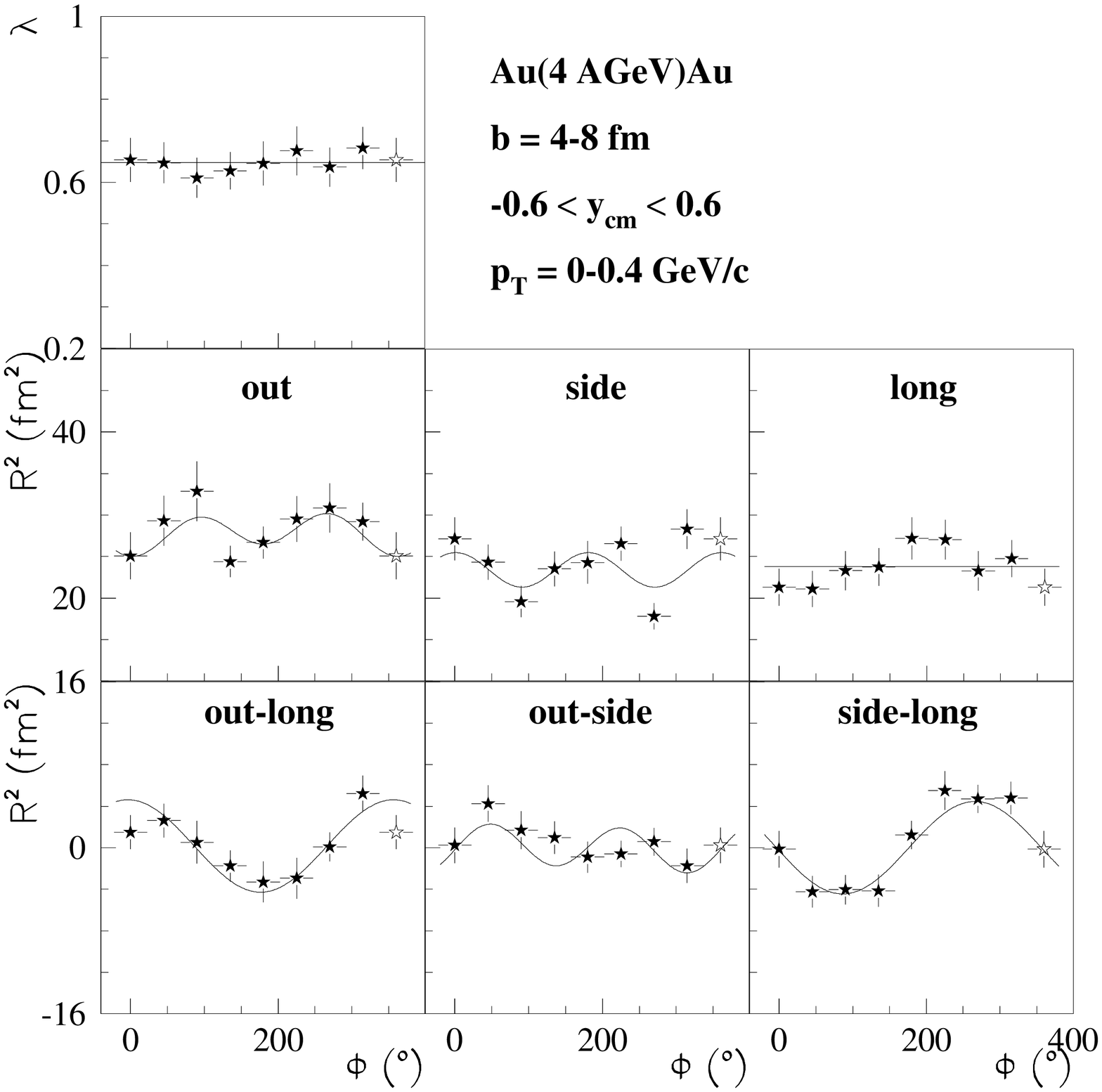,width=8cm}
\vspace*{-1cm}
\caption{Azimuthal dependence of
 Bertsch-Pratt 
HBT parameters
for midrapidity $\pi^-$ from peripheral Au+Au collisions
at 4~AGeV.  Lines represent a simultaneous fit to extract
the underlying geometry.}
\label{fig:4gev-radii}
\end{minipage}
\hspace{\fill}
\begin{minipage}[t]{70mm}
\epsfig{file=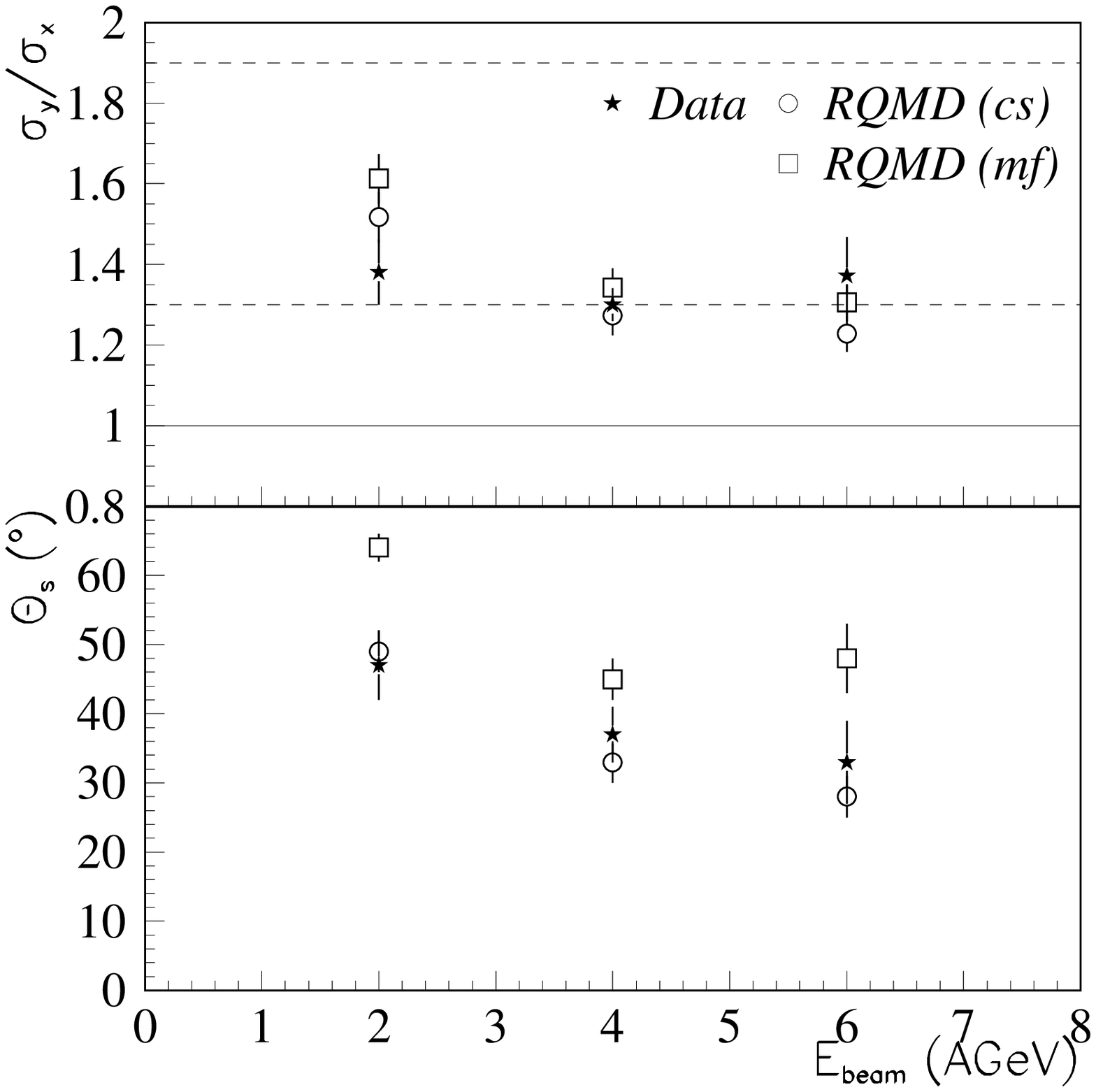,width=8cm}
\vspace*{-1cm}
\caption{
Excitation functions of the spatial analogs of elliptic (top)
and directed flow (bottom) for data (filled stars) and RQMD (open symbols) with and without meanfield.
}
\label{fig:tilt-summary}
\end{minipage}
\vspace*{-8mm}
\end{figure}

For 4~AGeV collisions, the seven fit parameters $R_{ij}^2$ and $\lambda$ are plotted 
as a function of $\phi$ in
Figure~\ref{fig:4gev-radii}.
The $\phi$-independence of $\lambda$ indicates that the fraction of $\pi^-$ from long-lived
resonances is independent of emission angle with respect to the reaction plane, consistent
with the relatively weak anisotropic flow at these energies~\cite{E895-v2},
as well as the randomizing effect on the pion from decay of the parent.
$R_{o}$ and $R_{s}$ show significant
equal and opposite
$2^{nd}$-order oscillations in $\phi$, while 
$R_{ol}^2$ and $R_{sl}^2$ display equal-magnitude $1^{st}$-order oscillations.


From the azimuthal dependence of the HBT radii, one may extract the geometry of the source in the 
{\it impact-parameter-fixed} coordinate system~\cite{E895-HBT2,LHW00}.  In particular, at midrapidity
and low $p_T$, $R_{ij}^2(\phi)$ are sensitive to the homogeneity lengths in x, y, z, and time, as
well as the tilt of the source in the reaction plane.
Thus, we determine the shape and orientation
of the freeze-out ellipsoid in {\it coordinate space}.
The lines in Figure~\ref{fig:4gev-radii} represent a simultaneous fit to all six $R_{it}^2(\phi)$,
using the formalism of~\cite{LHW00},
resulting in the following homogeneity lengths:
$\sigma_x=4.0\pm 0.1$ fm, 
$\sigma_y=5.2\pm 0.1$ fm, 
$\sigma_z=5.2\pm 0.1$ fm, 
$\sigma_t=3.7\pm 0.5$ fm/c; the extracted tilt angle is $\theta_s=37^\circ\pm4^\circ$.
Finite reaction-plane resolution effects, which wash out oscillations in $R_{ij}^2(\phi)$, are
accounted for~\cite{E895-HBT2}.

Similar shapes and orientations were found for collisions at 2 and 6~AGeV.
At all energies, our results indicate a pion freezeout distribution as an ellipsoid
whose major axis in the reaction plane is tilted with respect to the beam in the positive direction
(i.e. in the direction of ${\bf b}$),
and whose transverse axis perpendicular to the reaction plane is longer than the
axis in the reaction plane.  The extension in the temporal direction is consistent with
observations at high energy~\cite{E895-HBT1,NA44}.

The coordinate-space analog of elliptical flow, $\sigma_y/\sigma_x$, is shown as a function of
beam energy in Figure~\ref{fig:tilt-summary}.  In contrast to the momentum space trends, there
is no transition from out-of-plane to in-plane shape.
Indeed, the observed transverse shape is reminiscent of the entrance channel geometry.
In the simplest picture, pions are emitted from the overlap region of the two 
spherical Au nuclei.  For impact parameter $b=4-8$~fm, the ratio of out-of-plane to in-plane
RMS for this overlap region varies from 1.3-1.9 (indicated by dashed lines in Figure~\ref{fig:tilt-summary}),
consistent with the measured value $\sim 1.3$.

The tilt angle, shown in the lower panel of Figure~\ref{fig:tilt-summary},
is the coordinate-space analog of directed flow, and
is a geometric feature of the collision dynamics~\cite{LHW00}.
Studies with the RQMD model indicate that most of the low-$p_T$ pions arise from
$\Delta$ decay, so it is not surprising that the tilted freeze-out distribution resembles the baryonic
distribution in coordinate space calculated in transport codes~\cite{frankfurt-hydro}.


Experimental access to this level of geometric detail on the freezeout distribution
is unprecedented, and represents an exciting new opportunity to study the dynamical response
of hot nuclear matter to compression.  
An azimuthally-sensitive interferometry analysis~\cite{E895-HBT2} on pions generated by the RQMD~\cite{RQMD} model
displays considerable sensitivity on the dynamical effect of the nuclear meanfield.
Although at these energies the model underpredicts the spatial and overpredicts the temporal
scale (noted already for central collisions~\cite{E895-HBT1}),
qualitatively, the transverse shape and large positive
tilt angles are reproduced remarkably well; transverse
asymmetries and spatial tilt angles are shown in Figure~\ref{fig:tilt-summary}.
Since RQMD better describes proton directed flow (momentum-space tilt)
when the meanfield is included
in the calculation~\cite{E895-v1},
it is interesting to note that the spatial tilt angles
are better described when the meanfield is off.
Clearly, probing asymmetries in both momentum and coordinate space results in a more stringent test of the dynamics.

These coordinate-space anisotropies represent new and independent information.
Mom\-entum-space tilt angles (flow angles~\cite{FlowAngle}) at these energies are only a few
degrees~\cite{E895-v1}, and, as mentioned above,
for pions point
in the {\it negative}
direction~\cite{LHW00}, opposite the coordinate-space tilt $\theta_s$.
Experimental information on the interplay between coordinate and momentum space anisotropies
should help resolve theoretical issues, such as the coexistence of flow and antiflow 
components~\cite{frankfurt-hydro}.

\vspace*{-3mm}
\section{Six-dimensional Pion Phasespace Density}
\label{sec-PSD}

\begin{figure}[t]
\epsfig{file=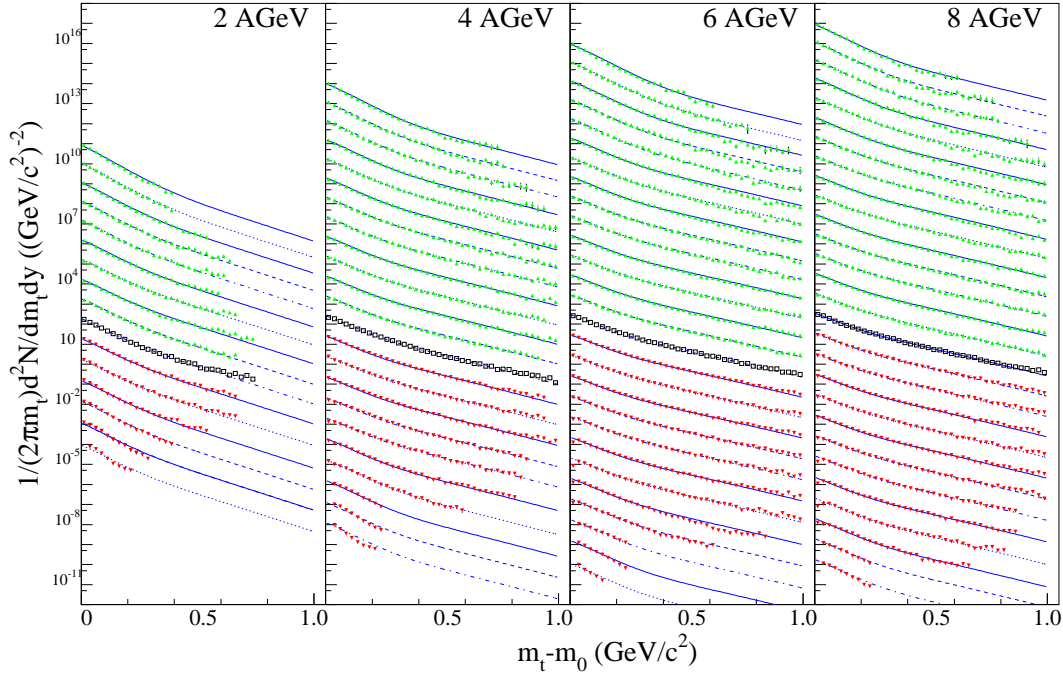,width=15cm}
\vspace*{-1cm}
\caption{Preliminary acceptance-corrected
 $\pi^-$ spectra from central Au+Au collisions at 2-8~AGeV, in 0.1-unit-wide rapidity slices.
The $m_T$ spectrum at midrapidity is shown by the open boxes; spectra for 
other rapidity slices are scaled by factors of ten.  Lines represent a two-component thermal fit.}
\label{fig:spectra}
\vspace*{-8mm}
\end{figure}

We have measured an excitation function of charged particle spectra over a wide region of phasespace.
Figure~\ref{fig:spectra} shows $m_T$ distributions for $\pi^-$ from central Au+Au collisions~\cite{E895-Klay}.
Assuming a thermal picture, these spectra may be combined with those of heavier particles, to
extract freezeout temperatures and radial flow values~\cite{E895-spectra}.

As a test of this thermal assumption, it has been observed~\cite{bertsch-PSD} that for a thermally equilibrated
pion source, both the shape {\it and} the amplitude of the six-dimensional phasespace occupation density as a
function of spatial and momentum coordinates, $f(x,p)$, are determined by the temperature.
  Experimentally, we 
may probe the spatially-averaged phasespace density by combining spectral and interferometry information,
according to~\cite{bertsch-PSD,ferenc-PSD}
\begin{equation}
\label{eq:PSD}
\langle f \rangle (p_T,y) = \sqrt{\lambda(p_T,y)\pi^3}\cdot
			   \frac{\frac{1}{E}\cdot\frac{d^3N}{dy\cdot p_Tdp_T\cdot d\phi}}
                    	         {R_o(p_T,y)\cdot R_s(p_T,y)\cdot R_l(p_T,y)} ,
\end{equation}
(For noncentral collisions or away from midrapidity, ``cross-term'' HBT radii also enter the denominator.)
The factor $\sqrt{\lambda}$ from the HBT analysis is included in Eq.~\ref{eq:PSD} to account for pions from long-lived
decays~\cite{WH99,ferenc-PSD}.

For heavy ion collisions at maximum AGS and SPS energies, a remarkable ``universal'' phasespace density
$\langle f \rangle (p_T,y)$ has been observed~\cite{ferenc-PSD}, independent of collision
energy or system, though multiplicities vary by more than an order of magnitude.
This suggests that $\pi-\pi$ cross-sections dominate the freeze-out density of 
high-energy collisions, and it is natural to look for a break-down of the universal behavior
at lower energy.

\begin{figure}[t]
\begin{minipage}[t]{80mm}
\epsfig{file=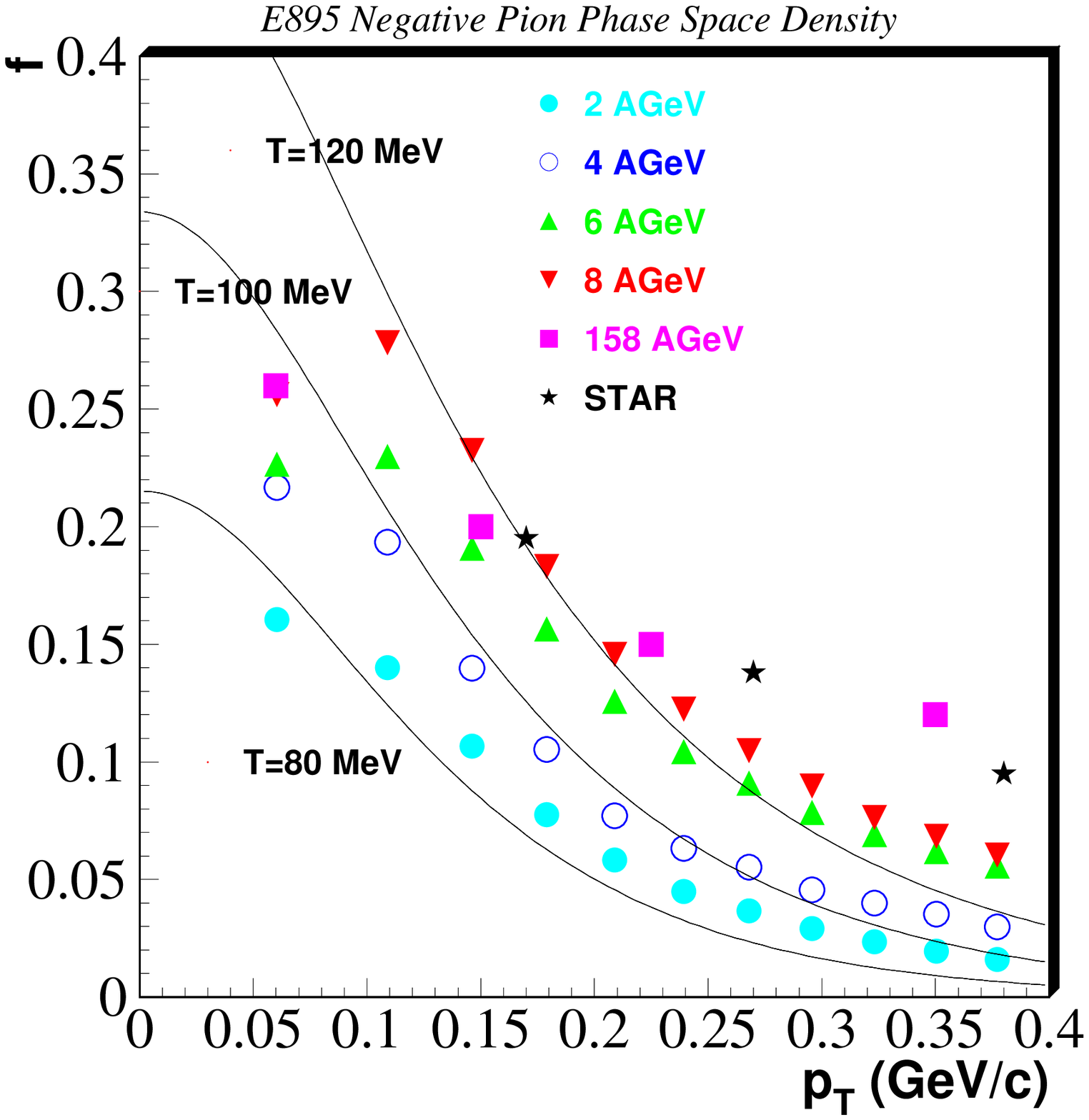,width=8cm}
\vspace*{-1cm}
\caption{
Preliminary phasespace density
 for midrapidity $\pi^-$ from 
central Au(Pb)+Au(Pb) collisions, measured by E895, NA49~\cite{ferenc-PSD}, and 
STAR~\cite{STAR-HBT}.
Uncertainties, not shown, are 10-20\%.
}
\label{fig:PSD}
\end{minipage}
\hspace{\fill}
\begin{minipage}[t]{75mm}
\epsfig{file=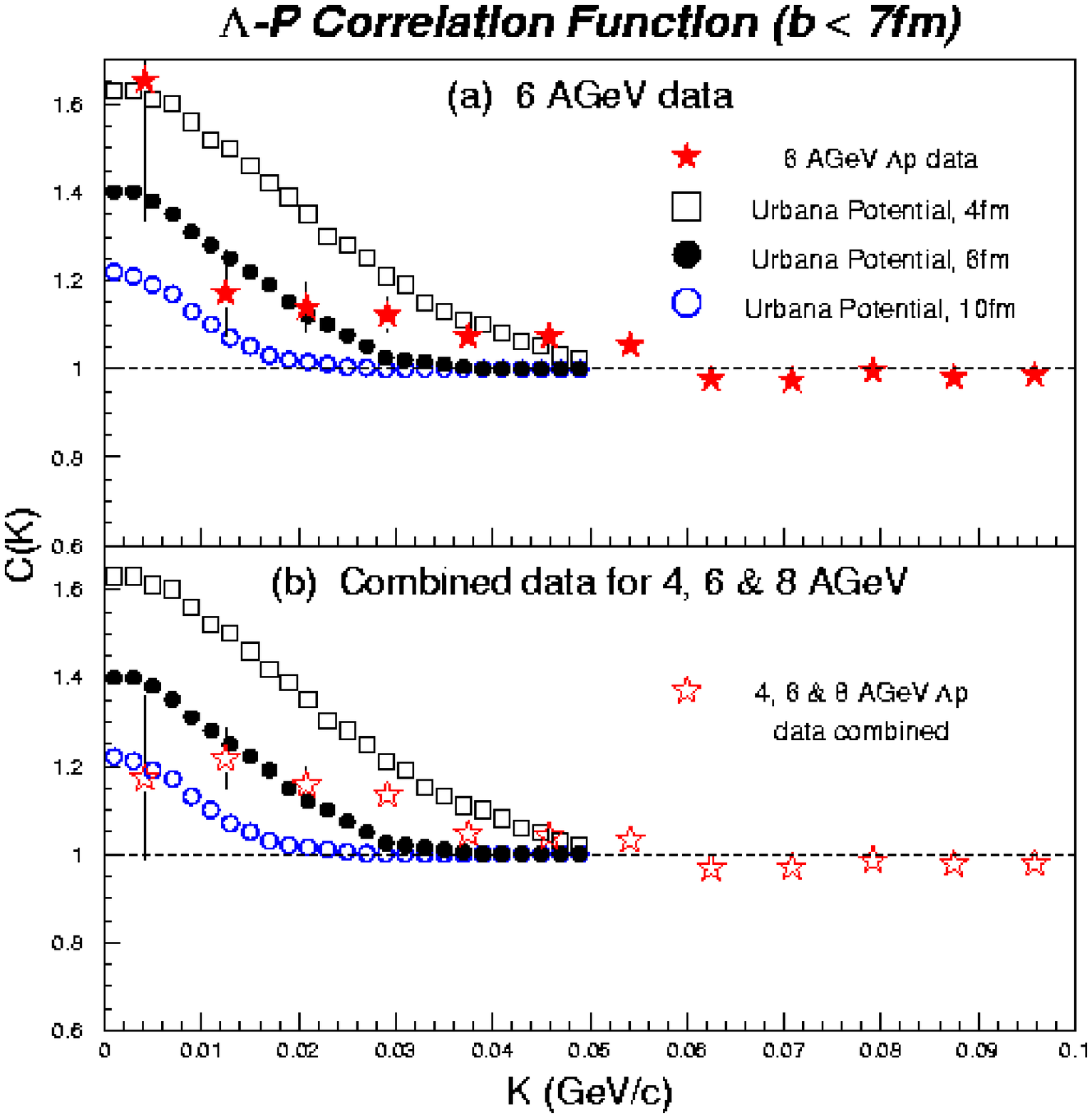,width=8cm}
\vspace*{-1cm}
\caption{Preliminary $\Lambda$-p correlation function
for
Au+Au collisions at 6~AGeV (a) and for combined
data from collisions at 2, 4, and 6~AGeV (b).
Also shown are expectations from a static source~\cite{wang-pratt}.
}
\label{fig:Lambda-p}
\end{minipage}
\vspace*{-8mm}
\end{figure}

Figure~\ref{fig:PSD} shows the $\pi^-$ 
average phasespace density $\langle f \rangle (p_T,y=0)$ determined according to Equation~\ref{eq:PSD}
for central Au+Au collisions at 2-8~AGeV.  Clearly, the universal behavior has
broken down; $\langle f \rangle$ grows monotonically with collision energy at the AGS.  As at higher energies,
$\langle f \rangle \ll 1$, indicating that multiparticle symmetrization effects are weak~\cite{WH99,ferenc-PSD}.
Lines indicate expectations
for ``pure thermal'' (no flow) sources with $T\sim 100$~MeV;
The extracted values appear roughly consistent with the shape and magnitude of a thermal source, with a
temperature increasing with $E_{beam}$.  
In detail, however, the observed $\langle f \rangle (p_T)$ is flatter than the ``pure thermal'' expectation,
consistent with a high-$p_T$ enhancement due to radial flow~\cite{ferenc-PSD}; preliminary fits (not shown) with a
parameterization that includes radial flow~\cite{tomasik-PhD} reproduce the distributions well, and indicate
that both temperature and flow increase with beam energy.

Also shown on Figure~\ref{fig:PSD} are values from Pb+Pb
collisions at 158~AGeV~\cite{ferenc-PSD} and preliminary data from STAR (Au(130 AGeV)+Au(130 AGeV))~\cite{STAR-HBT},
demonstrating the persistence of the universal behavior to RHIC energies.  While the phasespace densities achieved
at SPS and RHIC are higher than those at AGS at higher $p_T$, presumably due to the greater explosive flow generated,
it is remarkable that the ``universal'' value at low $p_T$ is reached already for collisions below maximum AGS energy.

\section{$\Lambda$-p Correlations}
\label{sec-pp-pLambda}

In the baryon-rich system generated at the AGS, it is interesting to study the
baryon freezeout geometry.  Two-proton correlations may be used to extract this
information~\cite{bauer_gelbke_pratt}; E895 proton interferometry results
indicate a proton freezeout volume independent of beam energy~\cite{E895-QM99-Rai}.
It has been suggested that $\Lambda-p$ correlation measurements could offer more sensitive
determinations of baryon source sizes than pp interferometry, particularly for large sources.
These measurements may also shed light on the $\Lambda$-p interaction, which is not fully
understood~\cite{wang-pratt}.

For semi-central ($b\le 7$ fm) Au+Au collisions, 
$\sim 120,000$ $\Lambda$s were identified in the TPC through
their decay topology into $p+\pi^-$,
using a neural network~\cite{E895-LambdaFlow,justice-v0}.
The $\Lambda$-p correlation function, as a function of
the relative momentum in the pair rest frame, is shown in Figure~\ref{fig:Lambda-p}.
A significant enhancement is seen at low relative momentum, consistent with
an attractive $\Lambda$-p final state interaction.

Also shown are expectations for the correlation function for various-sized static
sources, using a parameterized final state potential~\cite{wang-pratt}.
The shape of the experimental correlation function does not match these expectations
for any source size.  However, since 
the effects of the reconstructed $\Lambda$ purity ($\sim 80\%$)
and of $\Lambda$ feed-down from the electromagnetic decay of $\Sigma^0$s (estimated at $25-30\%$)
have not yet been accounted for, it is premature to draw conclusions regarding source
size or $\Lambda$-p potential.

\section{Summary}

The E895 Collaboration has performed
extensive studies of spectra, interferometry, strangeness, and flow at AGS energies,
providing valuable energy systematics useful for identification of ``new'' physics at SPS and RHIC.
Further, we have combined these analyses to extract new physics information.  
The azimuthal dependence of pion HBT radii reveals the anisotropic transverse shape
of the emitting source, and its tilted orientation in the reaction plane, imaging 
proton and pion flow in coordinate space for the first time.  The 6-dimensional $\pi^-$
phasespace occupation density is consistent with emission from a flowing, thermal source;
$\langle f \rangle_{\pi^-}$ grows with beam energy, approaching the ``universal'' limit observed
at SPS and RHIC energies.
First measurements of the $\Lambda$-p correlation function display an enhancement at low
relative momentum; more detailed study of this
correlation should provide valuable information on the baryon freezeout geometry and the
$\Lambda$-p potential.

\section*{Acknowledgements}
\vspace*{-3mm}
This work supported by the U.S. Department of Energy under contracts DE-AC03-76SF00098 and
DE-AC02-98CH10886, and grants DE-FG02-89ER40531, \linebreak DE-FG02-88ER40408, DE-FG02-87ER40324 and 
DE-FG02-87ER40331; by the U.S. National Science Foundation under grants PHY-98-04672, PHY-97-22653,
PHY-96-01271, PHY-96-05207, and INT-92-25096; by the University of Auckland Research Committee, 
NZ/USA Cooperative Science Programme CSP 95/33; and by the National
Natural Science Foundation of P.R. China under grant number 19875012.

\end{document}